%
%

	\documentclass[twocolumn,twoside]{IEEEtran}	

	\usepackage{xspace}
	\usepackage{bbold}
	\usepackage{citesort}
	\usepackage{amssymb}
	\usepackage{pstricks}
	\usepackage{theorem}
	\usepackage{times} 

	%
	\usepackage[final]{graphicx}
	\graphicspath{{./epsfig/}}	

	\input Def.tex	

	\def\KG{Ki\-ta\-ga\-wa \& Gersch\XS}
	
	\def\sss{\scriptscriptstyle}
	\def\LS{^{\sss \mathrm{LS}}}
	\def\RLS{^{\sss \mathrm{Reg}}}
	\def\KS{^{\sss \mathrm{KS}}}
	\def\S{^{\sss \mathrm{S}}}
	\def\ML{^{\sss \mathrm{ML}}}
	\def\Dk{\Delta_k}

	\newtheorem{remark}{Remark}
	\setlength{\itemindent}{0.25cm}

\bdoc

	\title{Regularized Adaptive Long Autoregressive \\ Spectral Analysis}

\author{
\JFG, \JI, Daniel \textsc{Muller} and Guy \textsc{Desodt}
\thanks{Jean-Fran\c{c}ois Giovannelli and J\'er\^ome Idier (\'emaux: giova@lss.supelec.fr, idier@lss.supelec.fr) are with the \ALSSun. Daniel Muller and Guy Desodt are with the Soci\'et\'e Thomson, 7 rue Mathurins, 92220 Bagneux, France.}
} 
\markboth
{\textit{T\lowercase{o appear in the }} IEEE Transaction on Geoscience and Remote Sensing}{Giovannelli \textit{\lowercase{et al.}}: Regularized Adaptive Long Autoregressive Spectral Analysis}

\maketitle

\begin{abstract} 
This paper is devoted to adaptive long autoregressive spectral analysis when ($\ib$) very few data are available, ($\ib\ib$) information does exist beforehand concerning the spectral smoothness and time continuity of the analyzed signals. The contribution is founded on two papers by \KG~\cite{Kitagawa85,Kitagawa85a}. The first one deals with spectral smoothness, in the regularization framework, while the second one is devoted to time continuity, in the Kalman formalism. The present paper proposes an original synthesis of the two contributions: a new regularized criterion is introduced that takes both information into account. The criterion is efficiently optimized by a Kalman smoother. One of the major features of the method is that it is entirely unsupervised: the problem of automatically adjusting the hyperparameters that balance data-based \vs prior-based information is solved by maximum likelihood. The improvement is quantified in the filed of meteorological radar.
\end{abstract}

\section{Introduction}

\PARstart{A}{daptive spectral analysis} and time-frequency a\-nalysis are of major importance in fields as widely varied as speech processing~\cite{Grenier86}, acoustical attenuation measurements~\cite{Kuc86,Idier94}, ultrasonic Doppler velocimetry~\cite{Peronneau91}, or Doppler radars~\cite{Barton97,LeFoll97,Dias00,Allan99,Barbaresco98}. Reference~\cite{Basseville92} gives a synthesis of the various methods for these problems, and provides a number of bibliographical introductions.

The present paper focuses on short-time analysis: typically, for analysis of pulsed Doppler signals only 8 or 16 samples are available to estimate one spectrum, with possibly various shapes (multimodal or not, of large spectral width or not, mixed clutter, \etc). Under such circumstances, the construction of the sought spectra becomes extremely tricky on the sole basis of the samples. 
As a point of reference, let us recall that several hundred samples are usually needed to compute an averaged periodogram with a fair bias-variance compromise~\cite{Allen77a,Allen77b}. So, parametric methods have generally been preferred, among which autoregressive (AR) play a central role. The AR coefficients estimation is usually tackled in the Least Squares (LS) framework~\cite{Marple87,Kay88}. These methods often provide a solution at points where non-parametric methods are useless; but when the number of data is very low, these techniques become, in their turn, useless, especially if various spectral shapes are expected due to model order limitations.

In order to construct a reliable image, structural information about the sought spectrum sequence must be accounted for. Our investigation is therefore restricted to the cases in which two kinds of information are foreknown: \textit{spectral smoothness} and \textit{time continuity}. This \aprio information is the foundation of the proposed construction. 

In the framework of stationary AR analysis, \KG proposed a method integrating the idea of spectral smoothness~\cite{Kitagawa85} by which a \textit{high-order AR model} can be robustly estimated, thereby getting around the difficult problem of order selection, and providing capability to estimate various spectral shapes. For the non-stationary case, and aside from~\cite{Kitagawa85}, the same authors introduced in~\cite{Kitagawa85a} a Markovian model for the regressor sequence in the Kalman formalism, in order to reflect time continuity. The present paper reviews~\cite{Kitagawa85} and~\cite{Kitagawa85a} and makes an original synthesis suited to the special configuration of Doppler signals. A new Regularized Least Squares (RegLS) criterion simultaneously includes the spectral and time information and is optimized by a Kalman Smoother (KS). 

One of the major features of the method is that it is entirely unsupervised: the adjustment of parameters that weight the relative contributions of the observation \versus the \aprio knowledge is automatically set by maximum likelihood (ML). 

A comparative study is proposed in the context of pulsed Doppler radars. Special attention is payed to atmospheric and/or meteorological context imaging or identification: ground clutter, rain clutter, sea echos,  \etc Adaptive spectral estimation of mixed clutter is achieved by means of several usual AR methods and the proposed one. The latter achieves qualitative and quantitative improvements \wrt usual methods. 

The paper is organized as follows. Section~\ref{Position} mainly introduces notations and problem statement. Section~\ref{Classique} focuses on usual LS methods and usual adaptive extensions. The proposed method is presented in Section~\ref{Methode} and Section~\ref{Kalman} deals with the KS. The problem of automatic parameter estimation is addressed in Section~\ref{Hyper}. Simulation results are presented in Section~\ref{Simul}. Finally, conclusions and perspectives for future works are presented in Section~\ref{Conclu}. 

\section{Problem statement}
\label{Position}

\begin{figure*}[htbp]
\cl{
\includegraphics[height=5cm]{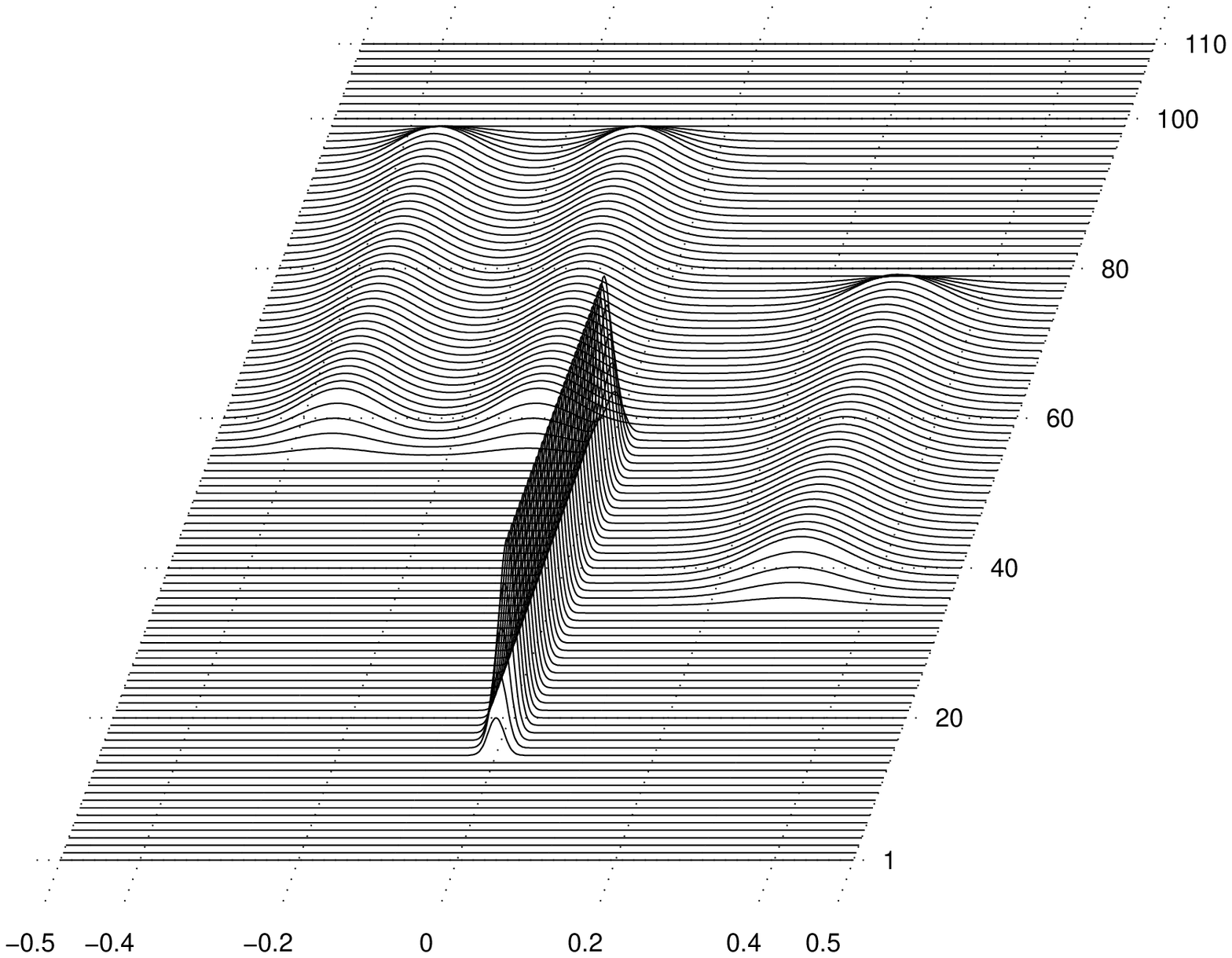}\hspace{-0.5cm}
\includegraphics[height=5cm]{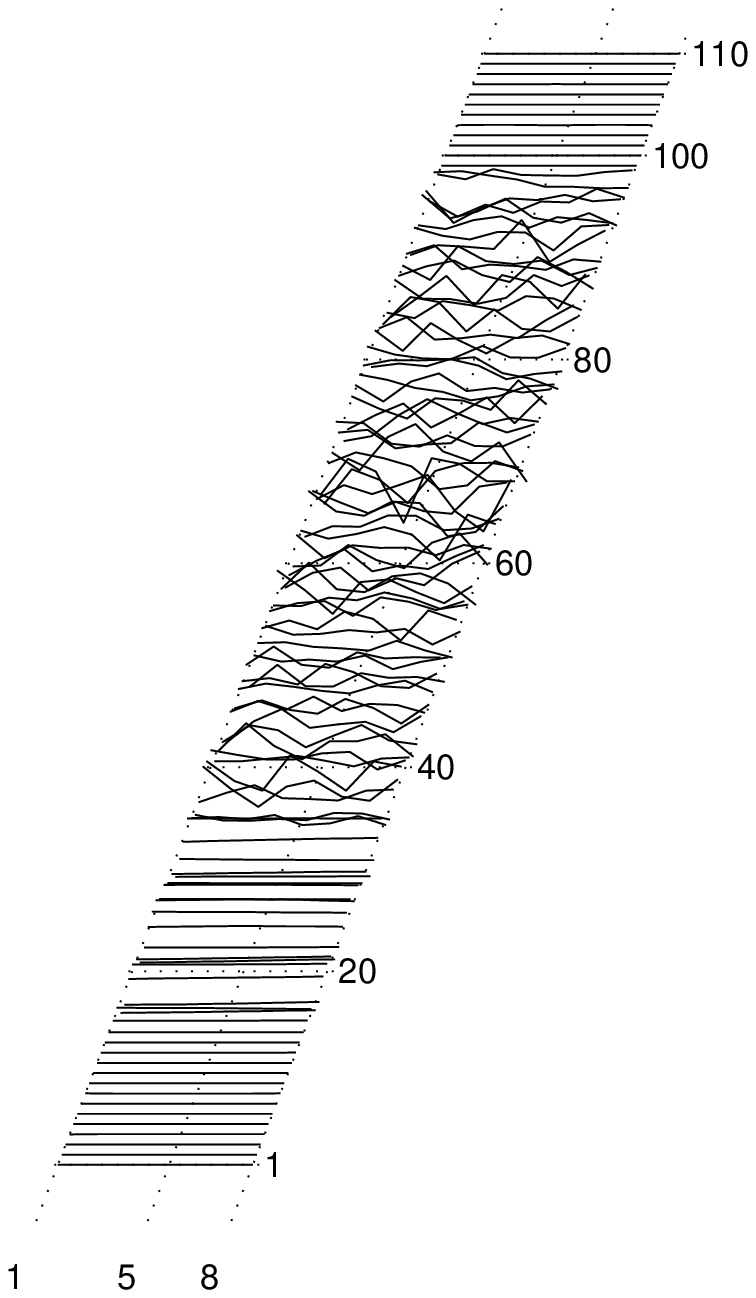}\hspace{-0.3cm}
\includegraphics[height=5cm]{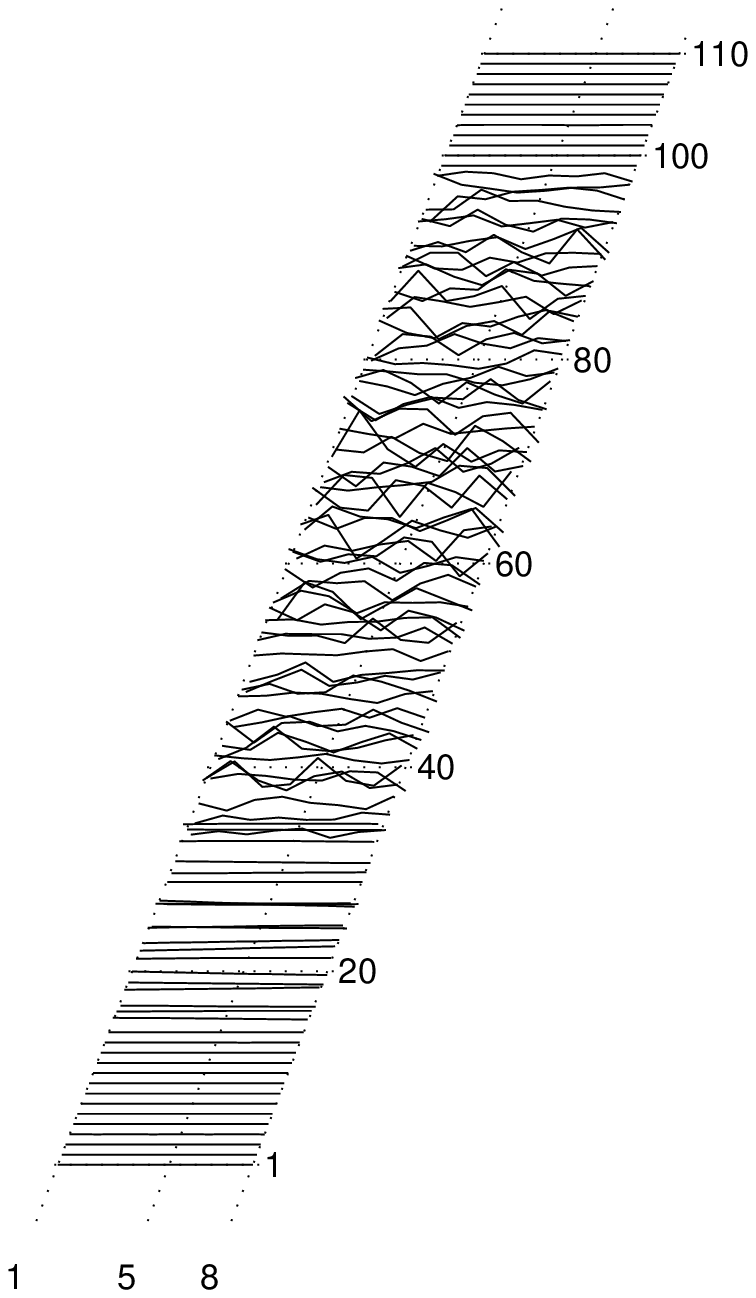}\hspace{-0.3cm}
\includegraphics[height=5cm]{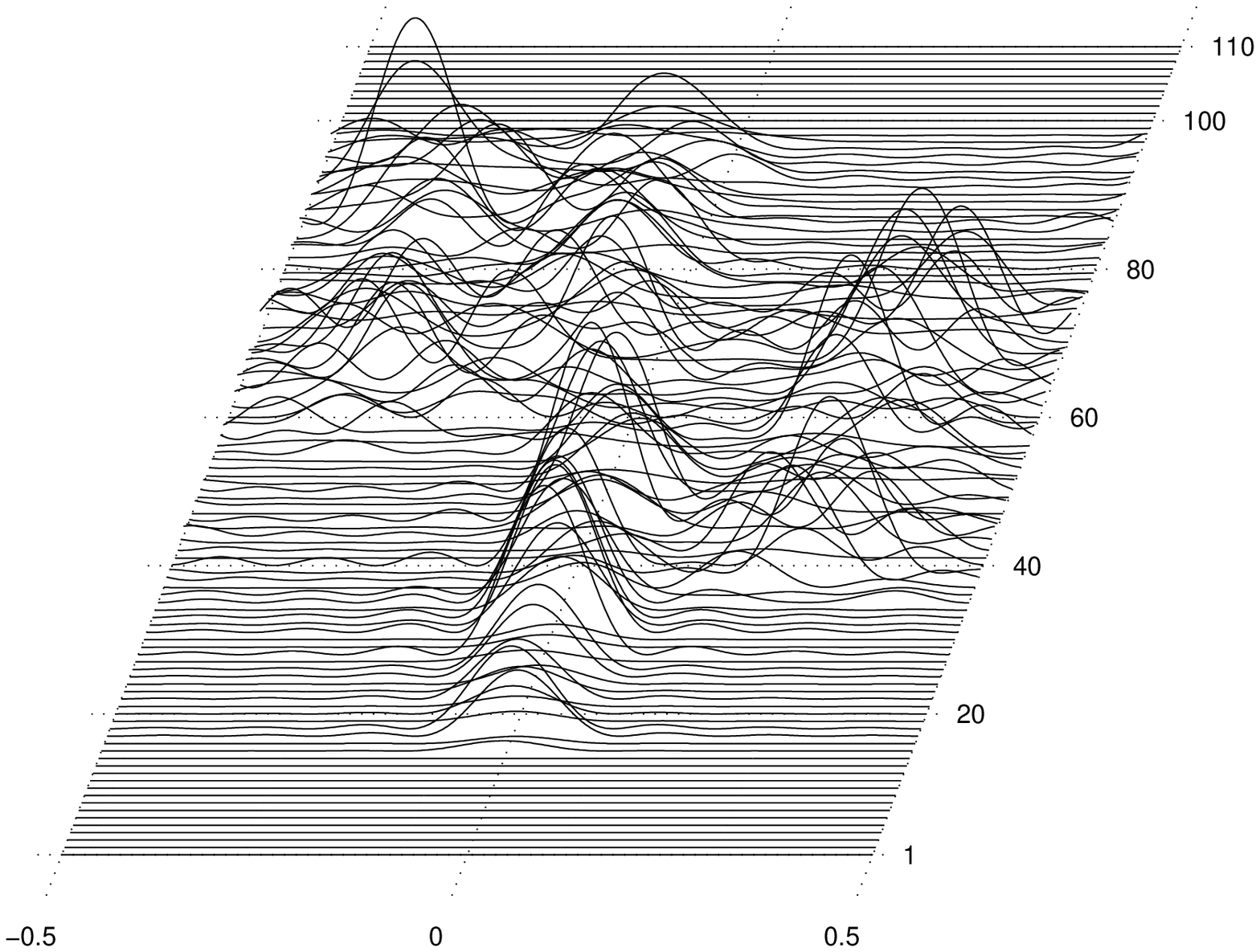}
}
\caption[Observed data and true spectra]{Simulated observations over 110 range bins with 8 samples per bin (corresponding to 8 Doppler pulses). The lhs figure shows the true spectra sequence. The narrow zero-mean spectra characterizes ground clutter (bin 15 to 57). Rain clutter induces more or less broad, single-mode spectra (bin 35 to 75). Lastly, sea echos resulting from wave phenomena exhibit two maxima (bin 56 to 95). The middle figure shows the real part and imaginary part of the data and the rhs one shows the associated periodograms.}
\label{FigObs}\label{VraiSpec}
\end{figure*}

The problem is that of processing pulsed Doppler signals from electronic scanning radars or ultrasound velocimeter. The reader may consult~\cite{Barton97,Peronneau91} for a technological review. The pulsed Doppler systems are such that the observed signals do not occur in the usual form of time-frequency problems. So, neither the usual time-frequency methods nor the one proposed by \KG can be directly applied, and part of the presented work consists in constructing an appropriate method for the encountered configuration.

The measurements are available as a set of complex signals $\Yc=[\yb_1, \dots, \yb_M]$, depth-wise juxtaposed in $M$ range bins. It is assumed that each $\yb_m=[y_{m1}, \dots, y_{mN}]^t$ is a $N$ sample vector extracted from a zero-mean stationary process. Fig.~\ref{FigObs} gives a Gaussian simulated example over $M=110$ bins for which $N=8$ samples are observed per bin. The successive regressors are denoted $\ab_m = \left[a_{mp}\right]$, where $m$ indicates the considered bin ($m\in\eN^*_M=\{1,2,\dots, M\}$) and $p$ the order of the autoregression coefficient ($p\in\eN^*_P$). Let us note $\Ab=[\ab_1, \dots, \ab_M]\in\eC^{N\times P}$ the collection of the whole set of coefficients. Let us also introduce $r_m$ and $r^e_m$ for signal and prediction error powers. The remainder of the paper is devoted to estimation of these quantities. The next section deals with the usual LS methods and their adaptive extension, and shows their inadequacy for the problem at stake.

\section{Review of classical methods} 
\label{Classique}

\subsection{Stationary spectral analysis}
\label{CasStatio}
\label{MCU}

This subsection is devoted to spectral analysis applied to a single bin $m$. Assuming a Gaussian distribution for the observed signal, the likelihood of the AR coefficients $\fxsp{\yb_m}{\ab_m}$ shows a special form~\cite[p.~82]{Picinbono91}, but its maximization raises a difficult problem. A few authors~\cite{Kay83,Pham88} have undertaken to solve it; but, firstly, the available algorithms cannot guarantee global maximization, and secondly, they are not computationally efficient for the applications under the scope of the paper. To remedy these disadvantages, the following approximation of the likelihood function is usually accepted~\cite[p.~185]{Kay88}: 
\beq \label{VraisAR}
 \fxsp{\yb_m}{\ab_m} = ( \pi r^e_m )^{-N} 
 \exp{ \left( - Q_m\LS (\ab_m)/r^e_m \right) } \,,
\eeq
involving the norm of the prediction error vector 
\beq \label{CritMC}
Q_m\LS (\ab_m)	=\eb_m^\dag \eb_m 
								= (\yb_m - Y_m \ab_m)^\dag (\yb_m - Y_m \ab_m) \,,
\eeq
\ie a quadratic form w.r.t. the $\ab_m$ namely the LS criterion. The $\yb_m$ and $Y_m$ are the vector and matrix designed according to some chosen windowing assumption~\cite[p.~217]{Marple87}, \cite[eq.~(2)]{Kay81}. There are four possible forms: non-windowed (covariance method), pre-windowed, post-windowed, double-windowed \ie pre- and post-windowed (autocorrelation method). Let us note $L$ the size of $\yb_m$: $L=N-P$, $L=N$ or $L=N+P$, according to the chosen form. This choice is of importance since it strongly influences spectral resolution for short time analysis~\cite[p.225]{Marple87}. 

Whatever the chosen form, the maximization of~(\ref{VraisAR}) comes down to the minimization of~(\ref{CritMC}) and yields:
\beq \label{SolMC}
\est{\ab}_m\LS 	= \argmin_{\ab_m} Q_m\LS(\ab_m ) 
						= ( Y_m^\dag Y_m)^{-1} Y_m^\dag \yb_m \,.
\eeq

As a prerequisite, the problem of choosing the model order $P$ must be tackled: $P$ has to be high enough to describe various PSD, and low enough to avoid spurious peaks, \ie to ensure spectral smoothness. This compromise can usually be set by means of criteria such as FPE~\cite{Akaike70}, AIC~\cite{Akaike74}, CAT~\cite{Parzen74}, or MDL~\cite{Rissanen78}, but, in the situation of prime interest here, they fail because the available amount of data is too small~\cite{Ulrych76}. Actually, there exists no satisfying compromise in term of model order, since too few data are available to estimate DSPs with possibly complex structures.

\subsection{Adaptive spectral analysis}
\label{CasAdapt}
\label{MCA}

For the ``multi range bin'' analysis, the first idea consists in processing each bin independently: according to the LS approach, it amounts to minimize a global LS criterion:
\beq \label{CritMCGlobal}
Q\LS (\Ab)	= \sum_{m=1}^M  Q_m\LS (\ab_m) \,.
\eeq
However, the resulting spectra hold unrealistic variations in the spatial direction (see Fig.~\ref{SpecEst}). In order to remedy this problem, the Adaptive Least Squares (ALS) approach accounts for spatial continuity by processing the data from several bins, possibly in weighted form, to estimate each $\ab_m$. A first approach uses a series of LS criteria including the data in a spatial window of length $W$. A widely used alternative is the exponential decay memory which uses geometrically weighted LS criteria, with parameter $\lambda\in[0,1]$. The latter is more popular because it is simpler: $\lambda$ is merely incorporated into a standard recursive LS algorithm~\cite[p.~266]{Marple87}. In both cases, the degree of adaptivity, \ie the spatial continuity is modulated by $W$ or $\lambda$. 



\subsection{Conclusion}

Whatever the variant, the main disadvantage of these approaches has to do with the parameter settings.

\bit

\item[--] From the spectral standpoint, smoothness is introduced in a roundabout fashion, \via the model order (adjusted by $P$) and the compromise no longer exists when the amount of data is reduced.

\item[--] From the spatial standpoint, continuity is also indirectly introduced (and tuned by $W$ or $\lambda$) and no automatic method for adjusting this parameters is available. 

\eit

These limitations are unavoidable in the simple LS formalism, and to alleviate this problem we resort to the regularization theory. In this framework, the proposed approach 
\bit

\item includes the spectral smoothness and spatial continuity in the estimation criterion itself;

\item allows long-AR model to be robustly estimated, and then various spectra to be identified;

\item provides automatic parameter setting, \ie an entirely unsupervised method.

\eit

\section{Long AR -- spatial continuity -- spectral smoothness}
\label{Methode}

\subsection{Spatial continuity model}
\label{DouxTemps}

The first idea consists in building a spectral distance. Following~\cite{Kitagawa85a}, starting with the PSD in bin $m$ 
\beq \label{SpecAR}
S_m(\nu)=\frac{r^e_m}{ \left\vert 1-A_m(\nu) \right\vert^{2} } \,,
						\,	A_m(\nu) = \sum_{p=1}^P a_{mp} ~ e^{-2 j \pi \nu p} \,,
\eeq
the proposed spectral distance between $S_m$ and $S_{m'}$ is founded on the $k$-th Sobolev distance between $A_m$ and $A_{m'}$: 
\beqx
	D_k(m,m') \propto \int_0^1 \left\vert 
	\frac{\dD^k}{\dD \nu^k} \left[A_m(\nu)-A_{m'}(\nu)\right] 
	\right\vert ^2 d\nu \,.
\eeqx
Calculations similar to those of~\cite{Kitagawa85a} yield a quadratic form: 
\beq \label{DefDouxTemp}
 D_k(m,m') = (\ab_m - \ab_{m'})^\dag \Dk (\ab_m - \ab_{m'}) \,,
\eeq
where $\Dk=\Diag{1^{2k}, \dots, P^{2k}}$ is the $k$-th spectral matrix. 

\subsection{Spectral smoothness model}
\label{DouxSpec}

The spectral smoothness measure proposed by \KG in~\cite{Kitagawa85a} (see also~\cite{Giovannelli96}), is easily deduced from~(\ref{DefDouxTemp}) as the distance to a constant DSP 
\beq \label{DefDouxSpec}
D_k(m) \propto \ab_m^\dag \Dk \ab_m \,.
\eeq
According to~\cite{Kitagawa85,Kitagawa85a}, $k\in\eZ_+$, but $\Dk$ as well as~(\ref{DefDouxTemp}) and~(\ref{DefDouxSpec}) can be extended to $k\in\eR_+$. 

\begin{remark}---
Strictly speaking, $D_k(m,m')$ and $D_k(m)$ are not spectral distances nor spectral smoothness measures since they are not functions of the PSD itself. However, they are quadratic and this has two advantages: it considerably simplifies regressor calculations (see Section~\ref{Kalman}) as well as regularization parameter estimation (see Section~\ref{Hyper}).
\end{remark}

\subsection{Double smoothness}
\label{MCR}

Starting with the spectral smoothness~(\ref{DefDouxSpec}) and the spatial distance~(\ref{DefDouxTemp}), a new quadratic penalization is introduced:
\beq \label{CritPriorGlobal}
Q^{\sss \infty}(\Ab)= \frac{1}{r_\sD}~ \sum_{m=1}^M D_k(m) 
					+ \frac{1}{r_\dD}~ \sum_{m=1}^{M-1} D_k(m,m+1) \,.
\eeq
It integrates both spectral smoothness and spatial continuity respectively tuned by $\lambda_\sD=1/r_\sD$ and $\lambda_\dD=1/r_\dD$. 

\begin{remark}---\label{Rem:Prior}
The penalization~(\ref{CritPriorGlobal}) has a Bayesian interpretation~\cite{Houacine90} as a Gaussian prior for the sought regressors:
\beq \label{Prior}
f(\Ab) \propto \Exp{-Q^{\sss \infty}(\Ab)}\,,
\eeq
useful for hyperparameter estimation, in Section~\ref{Hyper}.
\end{remark}

\subsection{Regularized least squares}

From the LS criteria~(\ref{CritMCGlobal}) and the penalization term~(\ref{CritPriorGlobal}), the proposed RegLS criterion reads:
\beqn \label{CritMCRDouble}
Q\RLS(\Ab)	&=& Q\LS(\Ab) + Q^{\sss \infty}(\Ab)\\
&=&\sum_{m=1}^M \frac{1}{r^e_m} (\yb_m-Y_m\ab_m)^\dag(\yb_m-Y_m\ab_m)\nonumber\\
&&	+~\frac{1}{r_\sD} \sum_{m=1}^M \ab_m^\dag \Dk \ab_m \nonumber \\
&&	+~\frac{1}{r_\dD} \sum_{m=1}^{M-1} 
	(\ab_m - \ab_{m+1})^\dag \Dk (\ab_m - \ab_{m+1}) \nonumber
\eeqn
involving three terms which respectively measure fidelity to the data, spectral smoothness and spatial regularity. The regularized solution is defined as the minimizer of~(\ref{CritMCRDouble}): 
\beq \label{SolRLS}
\est{\Ab}_{\RLS} = \argmin_{\Ab} Q\RLS(\Ab)
\eeq
\begin{remark}--- \label{Rem:RLS=Post}
The regularized criterion~(\ref{CritMCRDouble}) has a clear Bayesian interpretation~\cite{Houacine90}: likelihood~(\ref{VraisAR}) and prior~(\ref{Prior}) can be fused thanks to the Bayes rule, into a Gaussian \post law for the sought regressors:
\beq \label{Post}
f(\Ab\,|\,\Yc) \propto \Exp{-Q\RLS(\Ab)}\,,
\eeq
So, the solution~(\ref{SolRLS}) is also the MAP estimate.
\end{remark}

\subsection{Optimization stage}

Several options are available to compute~(\ref{SolRLS}). Since $Q\RLS(\Ab)$ is quadratic, $\est{\Ab}_{\RLS}$ is the solution of a $MP \times MP$ linear system. Moreover, since the involved matrix is sparse, direct inversion should be tractable but not recommendable here ($M=110$, $P=7$). Another approach may be found in gradient or relaxation methods~\cite{Bertsekas95} since $Q\RLS(\Ab)$ is differentiable and convex. But, given the depth-wise structure, another algorithm is preferred: Kalman Smoothing (KS). So, we resort to the initial viewpoint of \KG in~\cite{Kitagawa85a}. However, it is noticeable that~\cite{Kitagawa85a} does not mention the minimized criterion, where as our KS is designed to minimize~(\ref{CritMCRDouble}).

\section{Kalman smoothing}\label{Kalman}
\subsection{State-space form}

\bit

\item[--] The successive prediction vectors $\ab_m$ are related by a first-order state equation:
\beq \label{EqEtat}
\ab_{m+1}=\alpha_m \ab_m + \varepsilonb_m \,,
\eeq
in which each $\varepsilonb_m$ is a complex, zero-mean, circular, vector with covariance matrix $P^\varepsilon_m =r^\varepsilon_m\Dk \pmu$ and the $\varepsilonb_m$-sequence, is depth-wise white.

\item[--] The full state model also brings in the initial mean and covariance: the null vector and $P^a=r^a \Dk\pmu$, respectively.

\item[--] The observation equation is the recurrence equation for the AR model in each bin, written in compact form as
\beq \label{EqObs}
\yb_m = Y_m \ab_m + \eb_m \,, 
\eeq
\ie a generalized version of the one proposed in~\cite{Kitagawa85a}, adapted to depth-wise vectorial data. Each $\eb_m$ is a complex, zero-mean, circular, vector with covariance $r^e_m I_{L}$; the $\eb_m$-sequence, is also depth-wise white.

\eit

\begin{remark}---
\cite{Kitagawa85a} accounts for spatial continuity by means of a special case of Eq.~(\ref{EqEtat}): $\ab_{m+1} = \ab_m + \varepsilonb_m$. The latter has two drawbacks, though. Firstly, it is introduced apart from the idea of spectral smoothness. Secondly, from a Bayesian point of view, this equation is interpreted as Brownian process with an increasing variance, which may cause drifts to appear in the estimated spectra. On the contrary, the new coefficients $\alpha_m$ can be chosen in order to ensure stationarity of the model~(\ref{EqEtat}) or to minimize the homogeneous criterion~(\ref{CritMCRDouble}).
\end{remark}

\subsection{Equivalence between parameter settings} 

\subsubsection{Homogeneous criterion}

This section establishes the formal link between the parameters of the KS ($r^a$ and $\alpha_m, r_m^\varepsilon$) and those of the regularized criterion~(\ref{CritMCRDouble}) ($r_\dD$ and $r_\sD$). \cite[p.150--158]{Jazwinski70} states that the KS associated to~(\ref{EqEtat})-(\ref{EqObs}) minimizes: $Q\KS(\Ab)$
\beqn \label{CritLisseur}
&=& \sum_{m=1}^{M}\frac{1}{r^e_m}(\yb_m - Y_m \ab_m)^\dag(\yb_m - Y_m \ab_m) \nonumber\\
&+& \sum_{m=1}^{M-1} \frac{1}{r^\varepsilon_m} (\ab_{m+1}-\alpha_m \ab_m)^\dag \Dk\pmu 
																				(\ab_{m+1}-\alpha_m \ab_m)	\nonumber\\
&+& \frac{1}{r^a} \ab_1^\dag \Dk\pmu \ab_1 \,.
\eeqn
Partial expansions yield identification of~(\ref{CritMCRDouble}) and~(\ref{CritLisseur}) through the following count-down recursion.

\incirc{1} Initialization ($m=M-1$):
\beqx
\alpha_{M-1} 	= (1+\rho)\pmu \,, ~~\AND~
r^\varepsilon_{M-1} 			= r_\dD \alpha_{M-1} \,.
\eeqx

\incirc{2} Count-down recursion ($m=M-2, \dots, 1$):
\beqx
\alpha_m 	= (2+\rho-\alpha_{m+1})\pmu \,, ~~\AND~
r^\varepsilon_m 		= r_\dD \alpha_m \,.
\eeqx

\incirc{3} The last step yields the initial power:
\beqnx
r^a &=& r_\dD(1+\rho-\alpha_{1})\pmu\\
\eeqnx
with $\rho=r_\dD/r_\sD \in \eR_+^*$. These equations allow to precompute the coefficients of the KS in order to minimize~(\ref{CritMCRDouble}). 

\subsubsection{Limit model}

This section is devoted to the asymptotic behavior of the $\alpha_m$-sequence. For the sake of notational simplicity, the sequence is rewritten in a count-up form:
\beq \label{UpSeq}
\barr{lclcl}
m=1			&:~~&\tilde{\alpha}_{1}		&=& (1+\rho)\pmu \\
m\in \eN^*	&:~~&\tilde{\alpha}_{m+1}	&=& (2+\rho-\tilde{\alpha}_m)\pmu \\
\earr
\eeq
It is clear that $\tilde{\alpha}_1\in]0,1[$ since $\rho\in \eR_+^*$. Let us introduce $f(u)=(2+\rho-u)\pmu$. It is straightforward that $f(]0,1[) \subset ]0,1[$, so the entire $\tilde{\alpha}_m$-sequence remains in $]0,1[$. Moreover, if it exists, the limit $\alpha_{\sss \infty}\in[0,1]$ necessarily fulfills $f(\alpha_{\sss \infty})=\alpha_{\sss \infty}$. Elementary algebra yields: 
\beqn \label{AlphaLimit}
\alpha_{\sss \infty}&=&\left(\theta-\sqrt{\theta^2-4}\right)/2
\eeqn
with $\theta=2+\rho=2+r_\dD/r_\sD$. Finally, one can effortless see that $\forall u,v\in]0,1[$ we have $|f(u)-f(v)|\leqslant (1+\rho)^{-2}|u-v|$, \ie $f$ is a Lipschitz function with ratio in $]0,1[$. Hence, the sequence effectively converges towards $\alpha_{\sss \infty}$. It is also easy to see that the sequence is monotonous: increasing if $\alpha_1<\alpha_{\sss \infty}$ and decreasing otherwise. In the present case, comparison of $\alpha_1$ in~(\ref{UpSeq}) and $\alpha_{\sss \infty}$ in~(\ref{AlphaLimit}) shows that the $\tilde{\alpha}_m$-sequence is decreasing (in the count-up form), hence, $\alpha_m$ is increasing.

Finally, since $r^\varepsilon_m 	= r_\dD \alpha_m$, the corresponding limit state power is given by: 
\beq \label{VarEpsLimit}
r^\varepsilon_{\sss \infty} = r_\dD \alpha_{\sss \infty} \,.
\eeq

\subsubsection{Associated stationary criterion}

This section is devoted to the stationary limit model: the special case of Eq.~(\ref{EqEtat}), with $\alpha_m=\alpha_{\sss \infty}$ and $r^\varepsilon_m=r^\varepsilon_{\sss \infty}$, \ie a stationary first-order AR model for the $\ab_m$-sequence. The initial power is denoted $r^a_{\sss \infty}$ for notational coherence, even if it is not defined as a limit. It is actually defined according to $r^\varepsilon_{\sss \infty}$ and $\alpha_{\sss \infty}$ in order to ensure stationarity for the first-order AR model: $r^a_{\sss \infty}= r^\varepsilon_{\sss \infty}/(1-\alpha_{\sss \infty}^2)$.

Replacement of $\alpha_m, r^\varepsilon_m, r^a$ by $\alpha_{\sss \infty},r^\varepsilon_{\sss \infty},r^a_{\sss \infty}$ in Eq.~(\ref{CritLisseur}) yields the criterion minimized by the stationary~KS:

\beqnx
Q\S(\Ab)	
&=&\sum_{m=1}^M \frac{1}{r^e_m} (\yb_m-Y_m\ab_m)^\dag(\yb_m-Y_m\ab_m) 																									\nonumber\\
&&	+~\frac{(1-\alpha_{\sss \infty})^2}{r^\varepsilon_{\sss \infty}} \sum_{m=1}^M \ab_m^\dag \Dk \ab_m 																						\nonumber \\
&&	+~\frac{\alpha_{\sss \infty}}{r^\varepsilon_{\sss \infty}} \sum_{m=1}^{M-1} 
	(\ab_m - \ab_{m+1})^\dag \Dk (\ab_m - \ab_{m+1}) 				\nonumber\\
&&	+~\frac{\alpha_{\sss \infty}(1-\alpha_{\sss \infty})}{r^\varepsilon_{\sss \infty}} 
	( \ab_1^\dag \Dk \ab_1 + \ab_M^\dag \Dk \ab_M )
\eeqnx
where superscript ``S'' stands for stationary. Since we have: $r_\dD = r^\varepsilon_{\sss \infty}/\alpha_{\sss \infty}$ from Eq.~(\ref{VarEpsLimit}) and $r_\sD = r^\varepsilon_{\sss \infty}/(1-\alpha_{\sss \infty})^2$ from Eq.~(\ref{AlphaLimit}), one can effortless see that: 
\beqx \label{CritLisstatio2}
Q\S(\Ab) = Q\RLS(\Ab)	+~\frac{\alpha_{\sss \infty}(1-\alpha_{\sss \infty})}{r^\varepsilon_{\sss \infty}} 
	( \ab_1^\dag \Dk \ab_1 + \ab_M^\dag \Dk \ab_M ) \,.
\eeqx
So, the stationary criterion $Q\S(\Ab)$ and the initial homogeneous one $Q\RLS(\Ab)$ are equal apart from the edge effects, \ie two terms regarding the first and last regressors. As a consequence, the minimizer of $Q\RLS(\Ab)$ and $Q\S(\Ab)$ are practically equivalent and the latter is preferred since it does not require precomputation of the $\alpha_m$ and $r^\varepsilon_m$. 

\subsection{Kalman smoother equations}

\bit

\item[$\bullet$] Initialization ($m=1$)
\beqn \label{DebutInitKalman}
 \ab_{1\mid 1}	& = & 0	\\
 P_{1\mid 1}	& = & r^a_{\sss\infty} \Dk\pmu
	\label{FinInitKalman}
\eeqn

\item[$\bullet$] Filtering phase (for $m=2,\ldots,M$)

	\bit 
	\item Prediction step
\beqn \label{DebutFiltreKalman}
\ab_{m\mid m-1}	& = & \alpha_{\sss \infty} \ab_{m-1\mid m-1} \\
P_{m\mid m-1}		& = & \alpha_{\sss\infty}^2 P_{m-1\mid m-1}+ r^\varepsilon_{\sss\infty} \Dk\pmu
\eeqn

	\item Correction step
\beqn 
K_m				& = & P_{m\mid m-1} Y_m^{\dag} \\
R_m				& = & r^e_m I_{L} +K_m^{\dag} Y_m \\
\eb_m				& = & \yb_m-Y_m \ab_{m\mid m-1} \\
\ab_{m\mid m}	& = & \ab_{m\mid m-1}+K_m R\pmu_m \eb_m \\
P_{m\mid m}		& = & P_{m\mid m-1}-K_m R\pmu_m K_m^{\dag} 
\label{FinFiltreKalman}
\eeqn
	\eit

\item[$\bullet$] Smoothing count-down phase (for $m=M-1,\ldots,1$)
\beqn \label{DebutLissageKalman}
Q_m		 &=&\alpha_{\sss\infty} P_{m\mid m} P_{m+1\mid m}\pmu\\
\ab_{m\mid M}&=&\ab_{m\mid m}+Q_m\left(\ab_{m+1\mid M}-\ab_{m+1\mid m}\right) \\
P_{m\mid M}&=&P_{m\mid m}+Q_m \left(P_{m+1\mid M}-P_{m+1\mid m}\right)Q^{\dag}_m
\eeqn

\eit \label{FinLissageKalman}

\subsection{Fast algorithm}

Fast algorithms used to take a primordial position in the past decades, especially for real-time computations. More specifically, for adap\-tive spectral analysis of ultrasound Doppler signal, the \textsc{Ma\-ras\-ca} algorithm~\cite{Houacine90} has been used in a real-time high-resolution velocimeter prototype. But, it has two drawbacks, resulting in a rigid spectral and spatial continuity tuning. On the one hand, it proceeds by blocks and incorporates spatial continuity by using the regressor of the current block as a prior mean for the next one; on the other hand, the fast version is developed only for the zero-order smoothness ($k=0$). 

To our knowledge, no fast algorithm exists for the KF in the configuration of interest, mainly because of the structures of the state equation and the smoothness matrix. However, fast algorithm may be developed on the basis of high-order displacement matrices~\cite{Sayed94}. More precisely, it is easy to see that the displacement matrix of order $2k+1$ (if integer) is null for $\Delta_k$. Taking advantage of this property may result in a fast version of the proposed algorithm. 

However, calculation time problems are now less crucial than they used. The standard KS algorithm only takes $0.36~s$\footnote{The proposed algorithm has been implemented using the computing environment \textit{matlab} on a Personal Computer, Pentium~III, with a 450~MHz CPU and 128 Mo of RAM.} to process the entire data set of Fig.~\ref{FigObs}, so, real time computations can probably be achieved. 
\section{Hyperparameters estimation}
\label{Hyper}

The estimated $\ab_m$-sequence and spectra sequence depend on $M+4$ hy\-per\-pa\-ra\-me\-ters: smoothness and AR orders $k$ and $P$, power sequence $r^e_m$, and two regularization parameters $\lambda_\sD$ and $\lambda_\dD$.

\subsection{Power parameters}

The $M$ parameters $r^e_m$ are needed by the proposed RegLS method as well as the LS and ALS procedures and the same empirical estimates will be used for all of them. In the criterion~(\ref{CritMCRDouble}), parameters $r^e_m$ only act as weighting coefficients, so that the successive terms are of equivalent weight. The proposed empirical technique replaces the prediction error powers $r^e_m$ by the signal powers $r_m$ themselves. A simple empirical estimate $\est{r}_m=\yb_m^\dag \yb_m / N$ could be used. However, since the estimation variance is high for $N=8$, in practice, a more efficient technique consists in smoothing the sequence $\est{r}_m$. Let us note that~\cite{Kitagawa85a} proposes a scheme which is equivalent in principle.

\subsection{Order parameters}\label{HyperOrder}

The proposed framework allows to estimate long AR models to describe various spectral shapes. Moreover, by choosing the maximal order $P=N-1$ we get rid of the difficult problem of model order selection. In fact, as expected and confirmed in Section~\ref{OrderSensitivity}, as long as $P$ is large enough it does not affect significantly the spectral shape. 

On the other hand, to our experience, the smoothness order $k$ does not affect the spectrum sequence provided that $k\neq0$. So, the smoothness order is \aprio tuned to $k=1$, \ie a first order derivative spectra penalization. Moreover, Section~\ref{OrderSensitivity} also provides a quantitative sensitivity study of the spectra sequence \wrt this parameter.

\subsection{Regularization parameters}

The problem of regularization parameter estimation within the proposed framework is a delicate one. It has been extensively studied and several techniques have been proposed and compared \cite{Golub79,Titterington85,Hall87,Thompson91,Fortier93,Giovannelli96}. The ML approach is often chosen within the Bayesian framework, mentioned in Remarks~\ref{Rem:Prior} and~\ref{Rem:RLS=Post}. The Gaussian likelihood function~(\ref{VraisAR}) and the Gaussian prior~(\ref{Prior}) together yield a Gaussian marginal law for the observed samples $f(\Yc\,;\lambda_\sD, \lambda_\dD)$, \ie the regularization parameter likelihood. The Hyperparameter-Co-Log-Likelihood (HCLL) is easily computed, for a given hyperparameter set, as a function of innovation vectors $\eb_m$ and  covariances $R_m$, \ie two of the KF sub-products: 
\beqx
HCLL(\lambda_\sD, \lambda_\dD) = \sum_{m=1}^M \ln \det R_m + \eb_m^{\dag} R_m\pmu \eb_m \,, 
\eeqx
ignoring constant coefficients. This expression is the generalization of a more conventional identity, available for scalar observations~\cite{Kitagawa85a}. The error covariance matrix $R_m$ is an $L \times L$ matrix, $L$ possibly ranging from $L=1$ to $L=N+P$, according to the windowing form and model order. Since $L=1$ is selected in the presented computations, no specific algorithm has been developed for inversion nor determinant calculations. 

The ML estimate:
\beq \label{HyperMV}
(\est{\lambda}_\sD\ML, \est{\lambda}_\dD\ML) = \argmin_{\lambda_\sD, \lambda_\dD} HCLL(\lambda_\sD, \lambda_\dD)
\eeq
can be computed by means of several algorithms: coordinate/gradient descent algorithm~\cite{Bertsekas95} or EM algorithms~\cite{Shumway82,Levinson82}, but none of them can ensure global optimization. Here, the optimization stage is  tackled by means of a coordinate descent algorithm with a golden section line search~\cite{Bertsekas95}. Since $HCLL$ is a function of two variables only, the optimization stage only requires about 10~$s^1$.

\section{Simulation results and comparisons}
\label{Simul}

The present section assesses the effectiveness of the proposed method, compared to the usual ones by processing the example shown in Fig.~\ref{FigObs}.

\subsection{Quantitative comparison criterion}
\label{DefQuantit}

Since the true spectrum sequence is known in the presented simulations, quantitative criteria are computable on the basis of distances between estimated spectra $\est{S}_m(\nu)$ and true ones $S_m(\nu)$, accumulated over the $M$ bins. Normalized distances:
\beqx
\LD^r = \frac{\sum_{m=1}^{M} \int_0^1 |\est{S}_m(\nu)-S_m(\nu)|^r \dD \nu}
						{\sum_{m=1}^{M} \int_0^1 |S_m(\nu)|^r \dD \nu} \,,
\eeqx
with $r=1$ and $r=2$ have been computed. The normalization is chosen so that a null estimated spectrum results in a 100\% error. Practically, the integrals are approximated by discrete summation over the frequency domain $\nu=q/Q, q\in\eN_{Q-1}$ with $Q=1024$.

\subsection{Tuning parameters}

\subsubsection{Usual methods}

Since no automatic parameters tuning is available for usual methods, these parameters have been chosen in order to produce the best $\LD^2$ distance. Moreover, we have checked that such a quantitative procedure finds itself in good agreement with the visual appreciation. 

\ben

\item[--] First of all, it is noticeable that, even for a short model, the non-windowed and pre-windowed methods systematically yield numerous spurious peaks. The best results have been obtained with the post-windowed form\footnote{A possible explanation for this rather counterintuitive fact, is that the post-windowed form is somewhat "self penalizing", \ie the corresponding criterion incorporates quadratic penalization terms: $\ab_m^\dag M \ab_m$, where $M$ only depends upon the data.} (double-windowed behaves similarly) so, the estimated spectra are of poor resolution~\cite[p.225]{Marple87}. 

\item[--] As expected, since the true spectra show up to three modes, the best results have been obtained with $P=3$ for both LS and ALS. 

\item[--] Finally, as far as the ALS method is concerned, $W=20$ has been selected. 

\een

\subsubsection{Regularized method}

\begin{figure}[htbp]
\cl{\includegraphics[width=4.5cm]{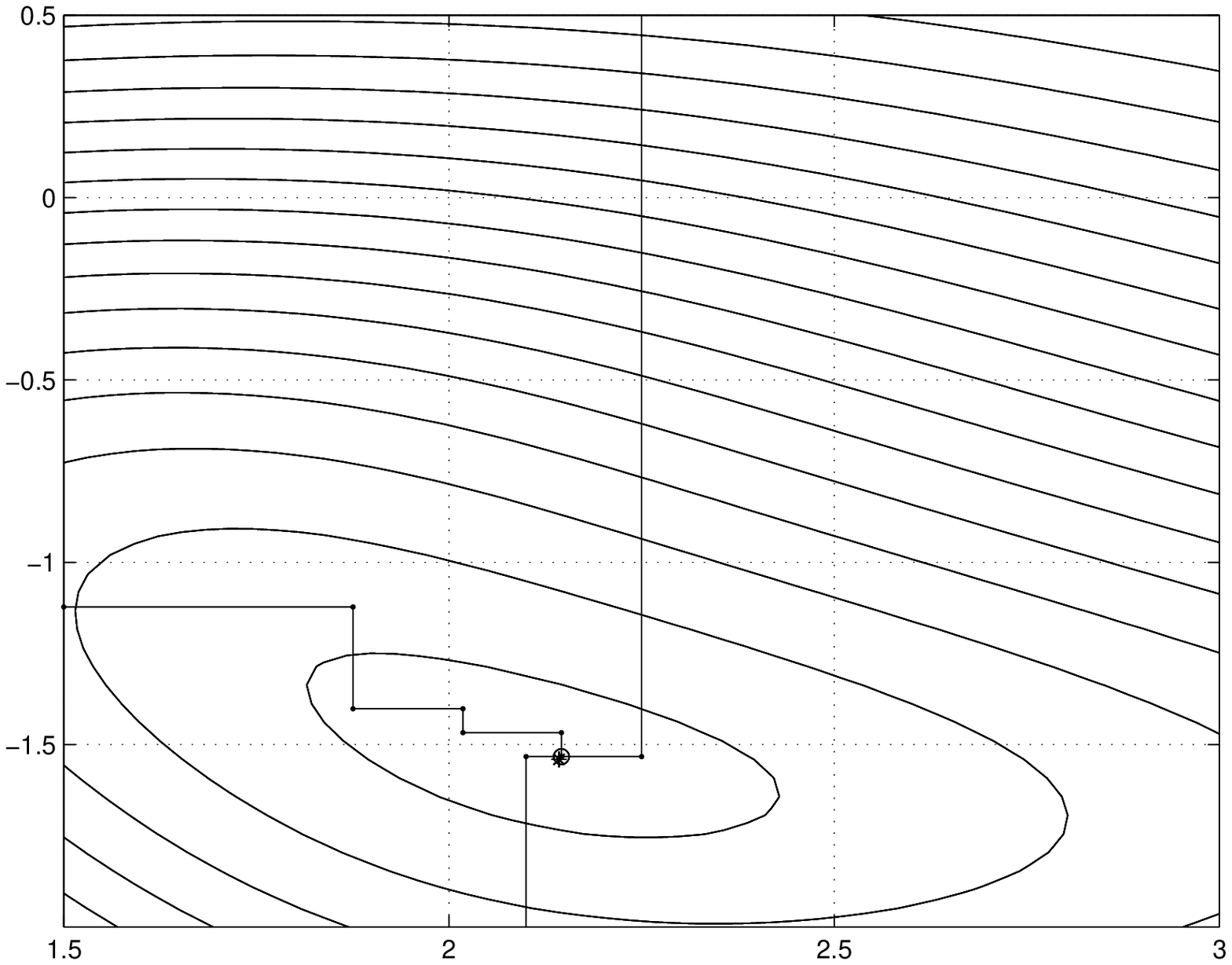} \includegraphics[width=4.5cm]{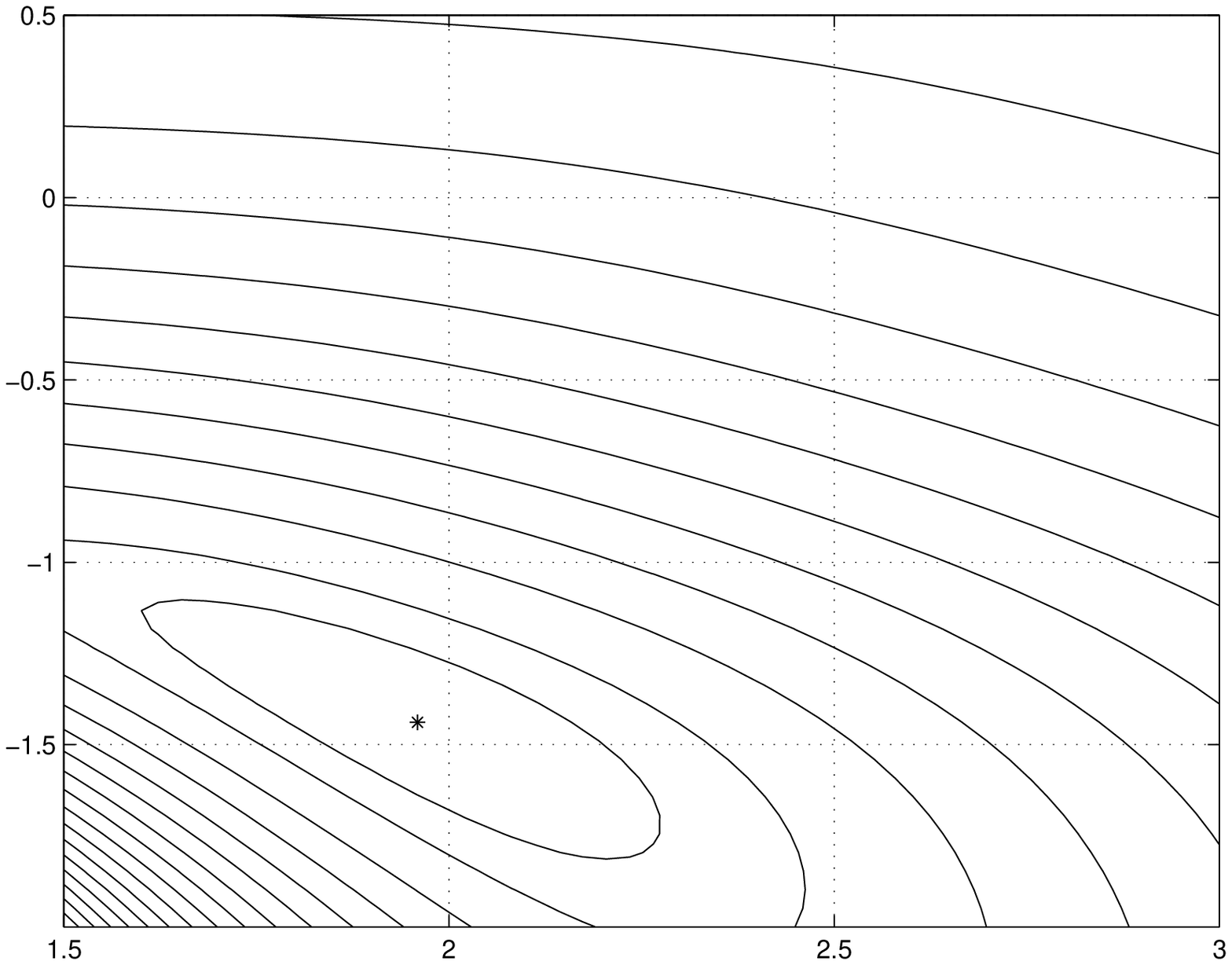}}
\caption[Sheet of co-log-likelihood and $\LD^2$ distance]{The lhs and rhs figure respectively show $HCLL$ and $\LD^2$ distance ($\LD^1$ behaves similarly) as a function of regularization parameters $(\lambda_\sD,\lambda_\dD)$, respectively read on the vertical and the horizontal axis ($\log_{10}$ scaled). In both cases, a star ($*$) locates the minimum.}
\label{FigVraisHyper}
\end{figure}

The HCLL function has been computed on a fine discrete $\log_{10}$ grid of $100\times 100$ values between ${-2}$ and ${1}$ for $\lambda_\sD$ and between ${1}$ and ${3}$ for $\lambda_\dD$. The result is the HCLL sheet shown in Fig.~\ref{FigVraisHyper}-lhs. It is fairly regular, and exhibits a single minimum at $\est{\lambda}_\sD\ML=-1.53$ and $\est{\lambda}_\dD\ML=2.16$. Moreover, Fig.~\ref{FigVraisHyper}-rhs shows the corresponding $\LD^2$ distances and the strikingly similar behavior of $HCLL(\lambda_\sD,\lambda_\dD)$ and $\LD^2(\lambda_\sD,\lambda_\dD)$ is a strong argument in favor of the likelihood as a criterion for parameters tuning. 

However, it must be mentioned that a variation of on decade on $\lambda_\sD$ or $\lambda_\dD$ entails a nearly imperceptible variation in the estimated spectra and a fraction of percent error. This point is especially important for qualifying the robustness of the proposed method. Contrary to the choice of model order in the usual AR analysis, which is critical, the choice of $(\lambda_\sD,\lambda_\dD)$ offers broad leeway and can be made reliably. 

Practically, the adjustment is set using the coordinate descent algorithm and Fig.~\ref{FigVraisHyper}-lhs illustrates its convergence, from three different starting points.

\subsection{Order sensitivity} \label{OrderSensitivity}

This subsection assesses the sensitivity of the method \wrt the order parameters $k$ and $P$. For $P=1$ to $P=7$ and for $k=0$ to $k=2$ (step .25), we have computed the ML estimate~(\ref{HyperMV}):
\beqx
(\est{\lambda}_\sD\ML(P,k), \est{\lambda}_\dD\ML(P,k)) 
= \argmin_{\lambda_\sD, \lambda_\dD} HCLL(\lambda_\sD, \lambda_\dD, P,k) \,,
\eeqx
and the corresponding optimal likelihood and distance 
\beqnx
HCLL_{\text{opt}}(P,k)	&=& HCLL(\est{\lambda}_\sD\ML(P,k), \est{\lambda}_\dD\ML(P,k), P,k) \\
\LD^r_{\text{opt}}(P,k)	&=& \LD^r(\est{\lambda}_\sD\ML(P,k), \est{\lambda}_\dD\ML(P,k), P,k)
\eeqnx
They are plotted in Fig.~\ref{FigVraisHyperKP}, as a function of $P$ for the several values of $k$. 

As far as the likelihood is concerned, 
\bit

\item $HCLL_{\text{opt}}$ is a decreasing (almost linear) function of model order $P$: the ML selected order is the maximal one $P=N-1=7$.  

\item $HCLL_{\text{opt}}$ does not depend on $k$ (the four curves are over plotted) so that, given $P$ the triplet $(\lambda_\sD,\lambda_\dD,k)$ ``over-parameterize'' the likelihood and $k$ is indifferent. 

\eit

As far as the $\LD^2$ is concerned, it still behaves similarly to the likelihood: it is roughly decreasing with $P$ and not depending upon $k$. As a conclusion, the maximization of the likelihood \wrt $k$ and $P$ does not provide any improvement and the recommended scheme described in Section~\ref{HyperOrder} is an efficient one.

\begin{figure}[htbp]

\cl{\includegraphics[width=7cm,height=2cm]{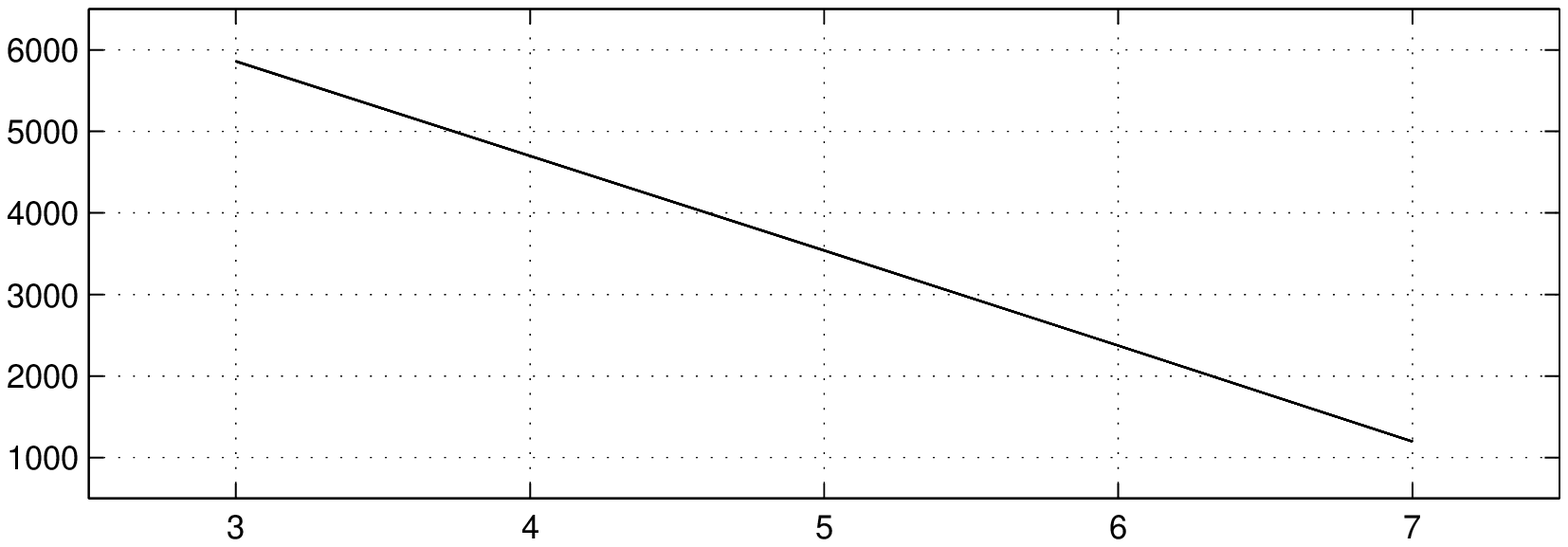}}
\cl{\includegraphics[width=7cm,height=2cm]{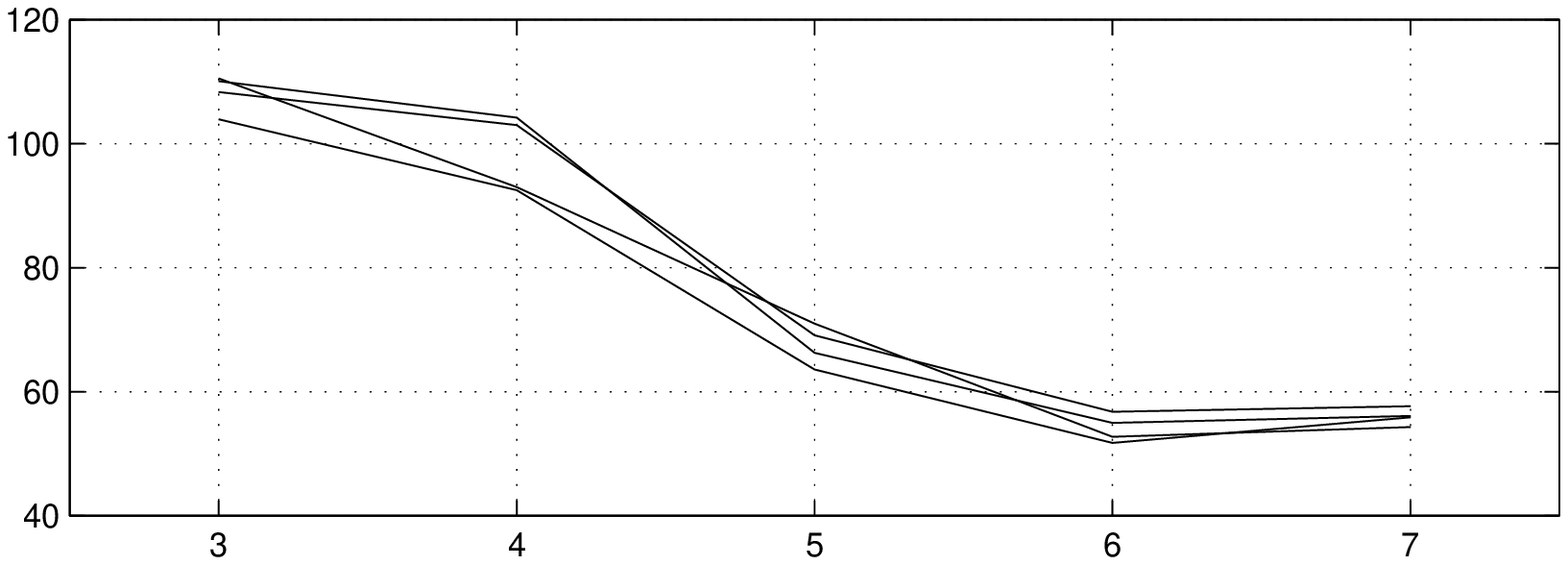}}
\caption[]{Optimal likelihood $HCLL_{\text{opt}}(P,k)$ (top) and distances $\LD^2_{\text{opt}}(P,k)$ (bottom) as a function of order $P$ for several smoothness order $k=0.5,1,1.5$, and $2$.}
\label{FigVraisHyperKP}
\end{figure}

\subsection{Qualitative evaluation}
\label{ResQualit}

\begin{figure*}[hbt]
\centerline{
\includegraphics[width=6cm]{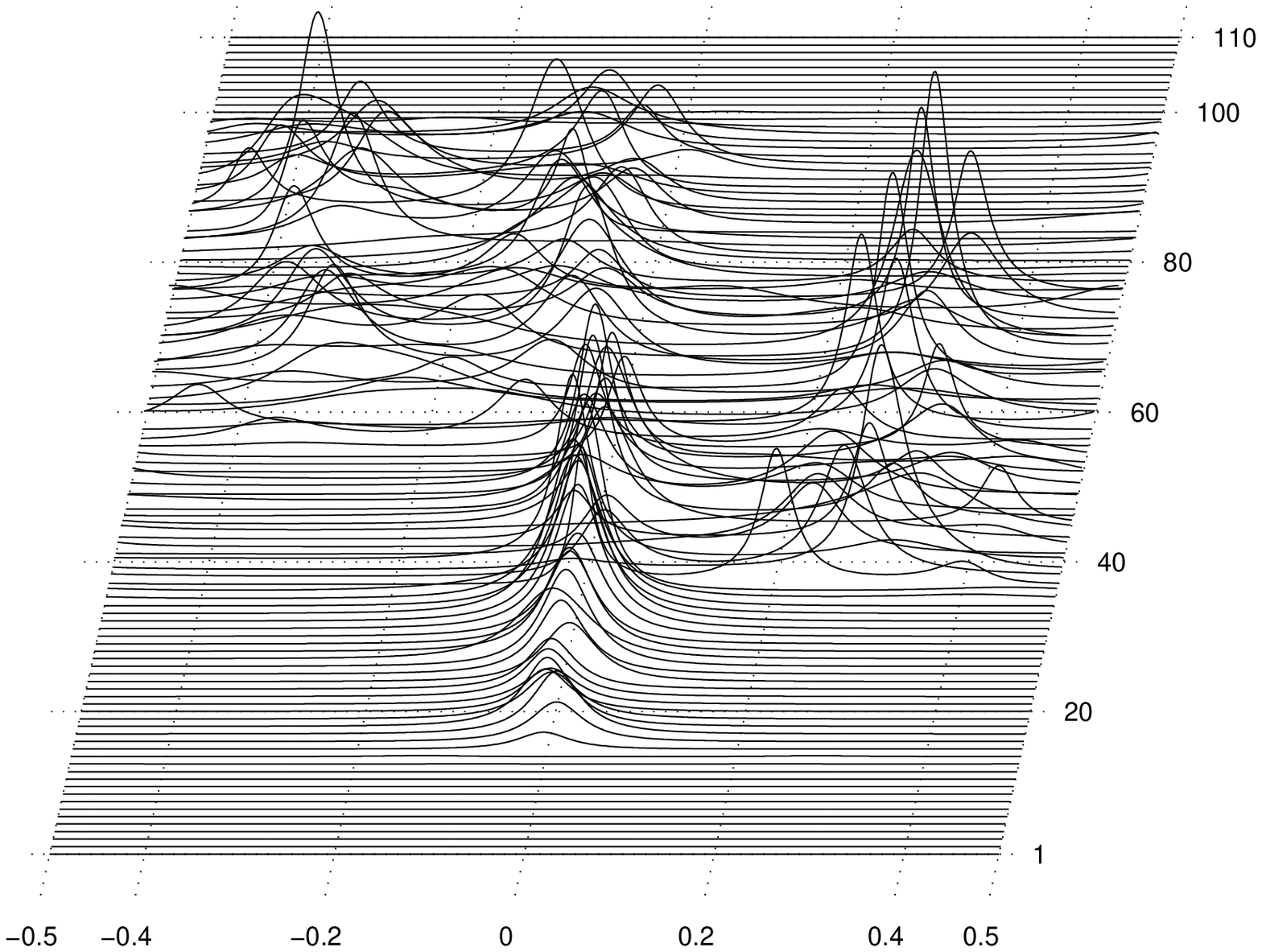}\hspace{-0.5cm}
\includegraphics[width=6cm]{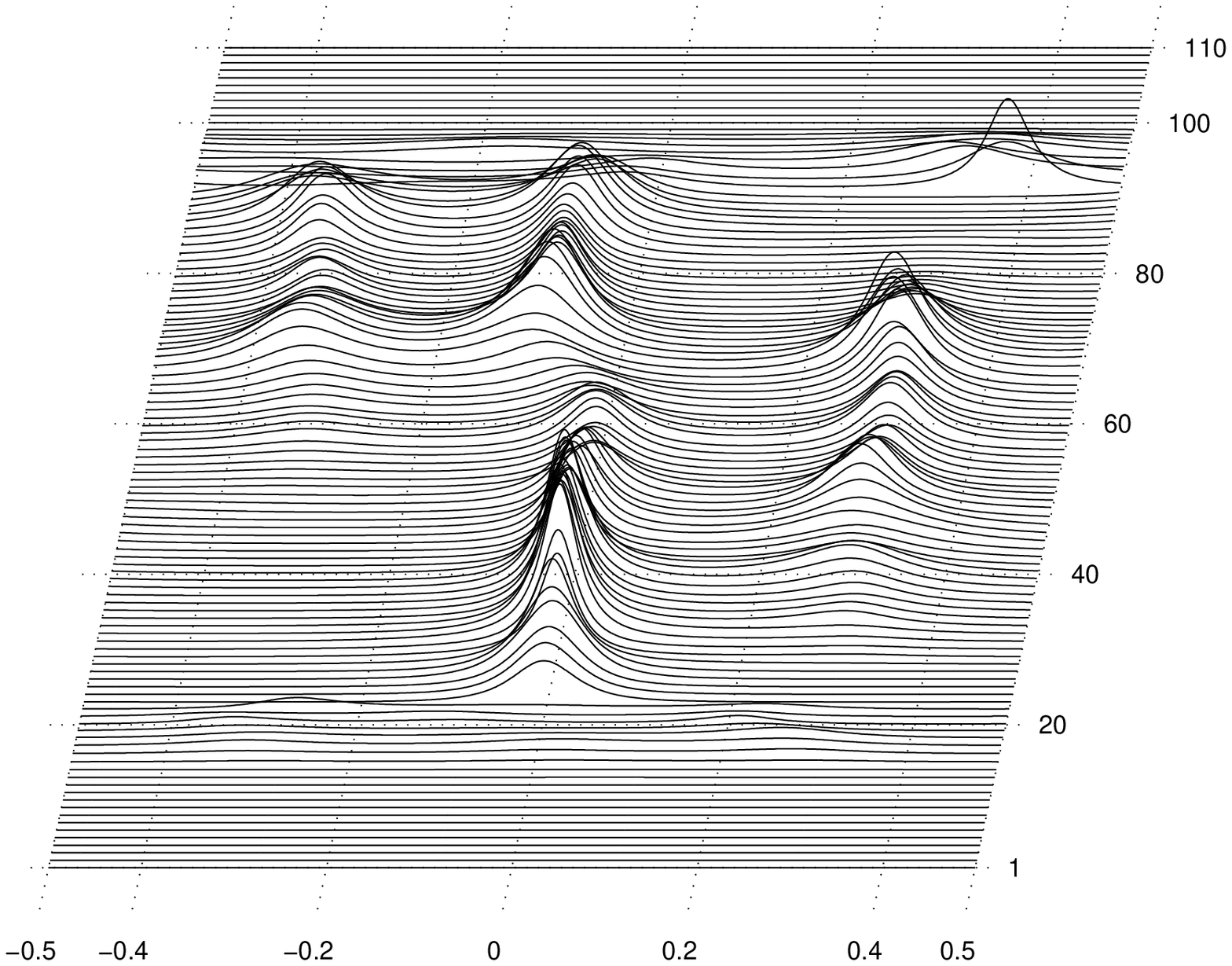}\hspace{-0.5cm}
\includegraphics[width=6cm]{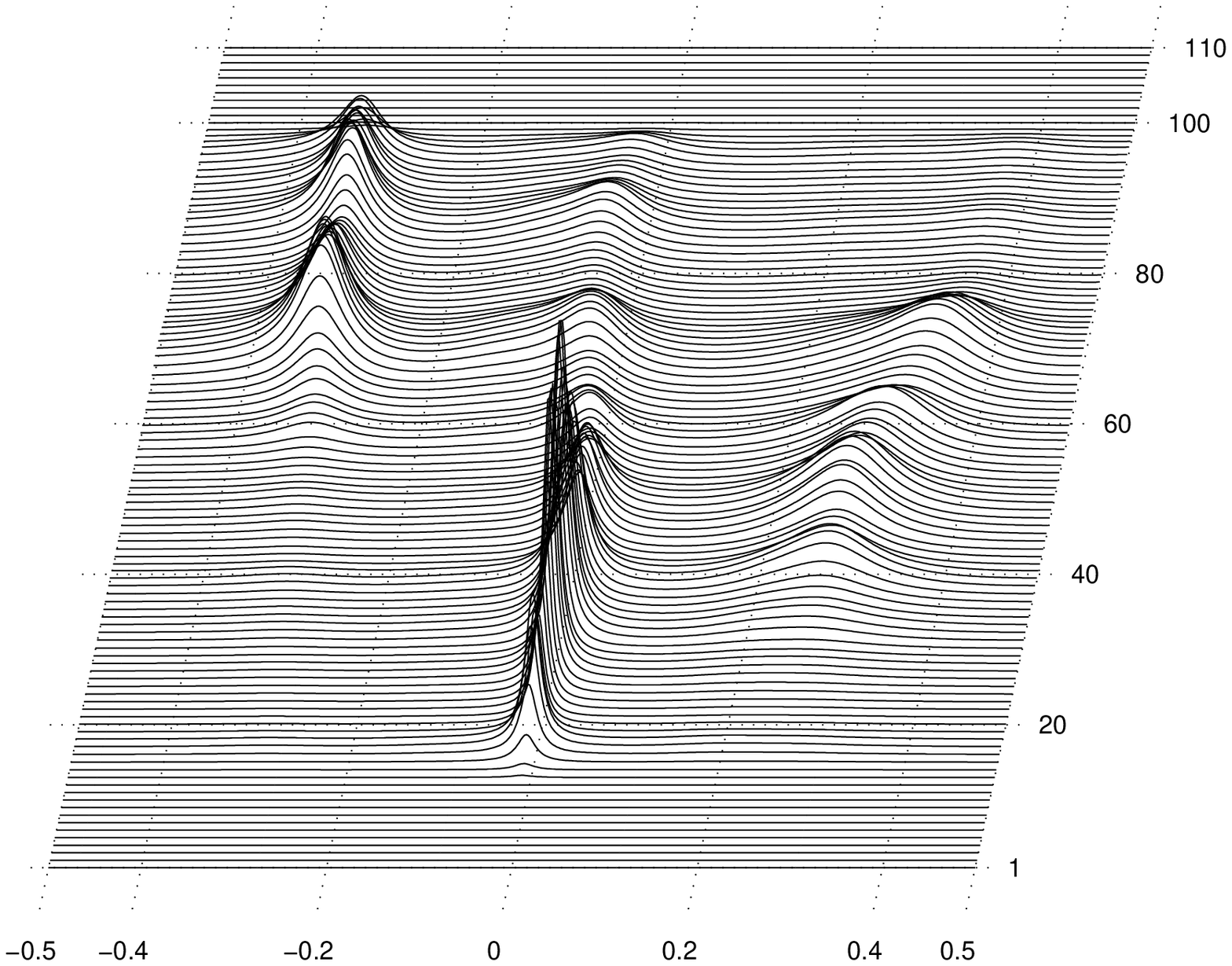}
}
 \caption[Estimated spectra]{Estimated spectra, from left to right: usual least squares estimate, adaptive least squares estimate and regularized least squares estimate (proposed method). Corresponding true spectra and data are shown in Fig.~\ref{FigObs}. Quantitative results are given in Table~\ref{Tab}.}
 \label{SpecEst}
\end{figure*}

We have then compared the usual methods at their best (optimally adjusted parameters knowing the true spectra) with the proposed method (automatic selection of regularization parameters without knowledge of the true spectra). The results obtained by LS, ALS, and RegLS are presented in Fig.~\ref{SpecEst}. A simple qualitative comparison with the reference Fig.~\ref{VraiSpec} already leads to four conclusions.
	
\bit

\item[--] The ML strategy provides a good value for the regularization parameters and the $\LD^2$ (and $\LD^1$) distance is in accordance with the qualitative assessment. 

\item[--] The effect of the regularization is obvious. Estimated spectra are in a much greater conformity with the true ones. The spectrum shapes are reproduced more precisely, in one, two or three modes. Their positions and their amplitudes are correctly estimated.

\item[--] Moreover, the spectral resolution for the ground clutter is strongly enhanced. It is essentially due to the coherent accounting for spectral and spatial continuity resulting in a robust non-windowed form. 

\item[--] However, it can be seen, though, that the sudden transitions at the beginning of the ground clutter is slightly over-smoothed. This can be expected from quadratic regularization and may be at least partially avoided by introducing non-quadratic regularization~\cite{Bouman93,Green90,Rudin92}. 
\eit

\begin{table}[hbt] 
\begin{center}
\begin{tabular}{|c|c|c|} \hline
Method		& $\LD^2$	& $\LD^1$	\\ \hline 
Periodogram	& 87.1\%		& 92.9\%		\\
Best LS		& 76.6\%		& 85.4\%		\\
Best ALS		& 66.4\%		& 75.5\%		\\
ML \& RegLS	& 57.9\%		& 69.2\%		\\ \hline
\end{tabular} 
\end{center} 
\caption[Quantitative comparison]{Quantitative comparison of the periodogram, least squares methods and the regularized one. $\LD^1$ and $\LD^2$ indicate the distances between estimated and true spectra.} 
\label{Tab} 
\end{table}

\subsection{Quantitative evaluation}
\label{ResQuantit}

In the non adaptive context, quantitative comparisons have previously been performed in~\cite{Kitagawa85,Giovannelli96}. The adaptive extension originally proposed by \KG has also been quantitatively assessed in~\cite{Kitagawa85a}.

For the proposed method, quantitative comparison have been achieved by evaluating $\LD^1$ and $\LD^2$ distances between true and estimated spectra. The results are listed in Table~\ref{Tab} and show an $\LD^2$ improvement of about 10\% form periodogram to best LS, 10\% from best LS to best ALS and 10\% from best ALS to the entirely automatic proposed method. 
\section{Conclusion and perspectives}
\label{Conclu}

This paper tackles short-time adaptive AR spectral estimation within the regularization framework. It proposes a new regularized least squares criterion accounting for spectral smoothness and spatial continuity. The criterion is efficiently optimized by a special Kalman smoother. In this sense, the present study significantly deepens the contributions of~\cite{Kitagawa85,Kitagawa85a}, given that the latter separately address spectral smoothness and spatial continuity. Moreover, the proposed method is entirely unsupervised and it is shown that maximum likelihood regularization parameters is both formally achievable and practically useful. Finally, a simulated comparison study is proposed in the field of Doppler radars. It shows an improvement of about 10\%, comparing some usual methods at their best \versus the entirely automatic proposed one. 

Future works will be devoted to compensate for the over-smoothing character of quadratic regularization in the presence of spatial breaks. \cite{Kitagawa87} accounts for spatial continuity while preserving breaks by way of a non-Gaussian state model and extended KF algorithms. In our mind, a preferable approach could be to introduce non-quadratic convex penalty terms and to minimize the resulting criterion using descent algorithms~\cite{Bouman93,Green90,Idier00b}. 
\section*{Acknowledgements}
Jean-Fran\c{c}ois Giovannelli is grateful to Mr Gr\"un and Mrs Groen for their 
expert editorial assistance.

\edoc